\documentclass[aps,pre,twocolumn,floatfix]{revtex4-1}
\usepackage[english]{babel}
\usepackage{amsmath, amsthm, amssymb, latexsym, graphicx}
\usepackage[utf8]{inputenc}
\usepackage[usenames,dvipsnames]{xcolor}
\usepackage[export]{adjustbox}
\usepackage[normalem]{ulem} %to strike the words

\begin{document}

\title{Interplay between Alfv\'en and magnetosonic waves in compressible
magnetohydrodynamics turbulence}
\author{N. Andr\'es$^{1}$}
\author{P. Clark di Leoni$^{2,3,5}$}
\author{P. D. Mininni$^{2,3}$}
\author{P. Dmitruk$^{2,3}$}
\author{F. Sahraoui$^{1}$}
\author{W. H. Matthaeus$^{4}$}

\thanks{Postprint version of the manuscript published in Physics of Plasmas {\bf 24}, 102314 (2017)}

\affiliation{$^1$ LPP, CNRS, Ecole Polytechnique, UPMC Univ. Paris 06, Univ. Paris-Sud, Observatoire de Paris, Université Paris-Saclay, Sorbonne Universités, PSL Research University, F-91128 Palaiseau, France\\ $^2$ Departamento de F\'{\i}sica, Facultad de Ciencias
Exactas y Naturales, Universidad de Buenos Aires, Ciudad Universitaria,
1428, Buenos Aires, Argentina.  \\ $^3$ Instituto de F\'{\i}sica de Buenos
Aires, CONICET-UBA, Ciudad Universitaria, 1428, Buenos Aires, Argentina.
\\ $^4$ Bartol Research Institute and Department of Physics and
Astronomy, University of Delaware, Newark, Delaware, USA. $^5$ Department of Physics and INFN, University of Rome “Tor Vergata,” Via della Ricerca Scientifica 1, 00133 Rome, Italy.}
\date{\today}

\begin{abstract}
Using spatio-temporal spectra we show direct evidence of excitation
of magnetosonic and Alfv\'en waves in three-dimensional compressible
magnetohydrodynamic turbulence at small Mach numbers. For the plasma
pressure dominated regime, or high $\beta$ regime (with $\beta$ the
ratio between fluid and magnetic pressure), and for the magnetic
pressure dominated regime, or low $\beta$ regime, we study magnetic
field fluctuations parallel and perpendicular to a guide magnetic
field \textbf{B}$_0$. In the low $\beta$ case we find excitation of
compressible and incompressible fluctuations, with a transfer of
energy towards Alfv\'enic modes and to a lesser extent towards
magnetosonic modes. In  particular, we find signatures of the presence
of fast magnetosonic waves in a scenario compatible with that of weak
turbulence. In the high $\beta$ case, fast and slow magnetosonic waves are present, with no clear trace of Alfv\'en waves, and a significant part of the energy is carried by two-dimensional turbulent eddies.
\end{abstract}

\maketitle

\section{Introduction}\label{intro}

Incompressible magnetohydrodynamics (IMHD) has a wide range of applications as a way to describe the large-scale behavior of different types of plasmas, including those of relevance for planetary science, astrophysics, and nuclear fusion science \citep{P1982,P1993,BiD1997}. However, this model is inadequate in those media where density fluctuations cannot be neglected. Examples of these environments are the ionized interstellar medium, some regions of the incoming solar wind, and planetary magnetosheaths \citep{A1981,B1982,BC2005}. For instance, recent in situ observations have shown that compressibility plays a significant role in the turbulent dynamics of the fast and slow solar wind, in particular  by supplying the energy dissipation needed to account for the local heating and particle acceleration of the solar wind \citep{C2009b,G2016,B2016c,H2017a,H2017b}. Thus, a study of compressible MHD (CMHD) turbulence is essential for a deep understanding of the turbulent dynamics of the solar wind at scales larger than the ion inertial length.

In presence of a uniform magnetic guide field \textbf{B}$_0$, the IMHD model has Alfv\'en waves as exact non-linear solutions. These transverse and incompressible waves propagate along the \textbf{B}$_0$ direction. When a turbulent regime develops in the presence of waves and eddies, two different regimes can be identified depending on the strength of the non-linear coupling, the so-called  weak and strong turbulent regimes. In IMHD, the strength of the nonlinear effects is related to the parameter $\chi=(k_\perp \delta B)/(k_\parallel B_0)$, i.e. the ratio between the nonlinear eddy turnover time $\tau_{nl}=k_\perp\delta v_\perp$ and the linear Alfv\'en time $\tau_A=k_\parallel u_A$. In the limit $\chi<<1$, the dynamics is controlled by weakly coupled waves, and perturbation theory can be used to obtain a prediction for the scaling of the energy spectrum \citep{Z1965,Ga2000,Ga2003,N2011}. When  $\chi \gtrsim1$, waves and eddies coexist with strong coupling, and phenomenological models are often used to study the nonlinear dynamics of turbulent plasmas \citep{I1963,K1965,H1984,G1995}. Note however that even in this case, some exact laws, e.g. the so-called 4/5 law of homogeneous turbulence, can be derived for different fluid approximations of magnetized plasmas \citep{vkh1938,Ch1951,P1998a,G2008,Ga2011,B2013,A2016a,A2016b,A2017}. It is important to recognize that, in general, the nonlinearity parameter $\chi$ may take on greatly differing values in different regions of ${\bf k}$-space.

The existence in IMHD of multiple time scales (the eddy turnover time, the Alfv\'en time, and the Alfv\'enic crossover time) gives rise to multiple phenomenological models of IMHD turbulence. In the so-called Iroshnikov-Kraichnan (IK) phenomenology \citep{I1963,K1965}, the interaction between waves and eddies results in a quenching of the energy transfer towards small scales, which are assumed to be isotropic. This results in a modification of the Kolmogorov energy spectrum from $E(k)\sim k^{-5/3}$ \citep{M1982,S2009,A2014a,A2016a,A2016b} to $E(k)\sim k^{-3/2}$ \citep{S1967,M1989,N2010,P2007,BC2013}. The anisotropy of IMHD turbulence has been extensively studied in the literature \citep {Mo1981,Sh1983,O1994,Mo1995,M1996,G1999,BN2001,A2011,MARV2011,S2016}. This has resulted in several phenomenological theories that drop the assumption of isotropy but in which the interactions between waves, and of waves with eddies, still play a central role \citep[see, e.g.,][]{H1984,G1995}.

Recently, the deep relation between waves and turbulence has been the subject of intensive research \citep{FL2007,DM2009,MD2011,Me2011,Me2016,L2016}. To identify the nature of waves in numerical simulations or experiments the spatio-temporal spectra have been widely used \citep{Me2015,Me2016,L2016,K2017}. Using direct numerical simulations of the IMHD equations with a uniform magnetic field, \citet{DM2009} focused on the properties of fluctuations in the frequency domain. The authors found the presence of peaks at the corresponding Alfv\'en wave frequencies in fully developed turbulent regimes, and nonlinear transfer of energy at wave numbers perpendicular to the mean magnetic field. \citet{Me2015} performed three dimensional (3D) numerical simulations of incompressible weak MHD turbulence and found evidence of accumulation of energy in Alfv\'en waves and in intermittent structures, while \citet{Me2016} investigated the transition of turbulence from weak to strong regime. \citet{L2016} considered relatively small, medium, and large values of the guide field ${\bf B}_0$ in IMHD simulations, and found that time decorrelation of Fourier modes is dominated by sweeping effects, and only at large values of \textbf{B}$_0$ and for wave vectors mainly aligned with this field time decorrelations are controlled by the Alfv\'enic time.

In comparison to IMHD turbulence, CMHD is more intricate due to nonlinear coupling of the velocity, magnetic field, density and pressure fluctuations \citep[see, e.g., ][]{Z1990,Z1992,Z1993}. In the CMHD approximation this emerges as the presence of two additional propagating wave modes that are not present in the IMHD model, namely fast and slow magnetosonic modes. These compressible modes can deeply affect the nonlinear dynamics of turbulent plasmas. Moreover, these modes or their counterparts in kinetic theory were reported using in situ spacecraft measurements in the solar wind \citep[see, e.g.][]{Kl2014,W2016,O2016}, planetary magnetosheath \citep{Sa2003,Sa2006,R2013,H2015,Hu2015,H2016} and foreshock regions \citep{Be2007,A2013,A2015}.

Different theoretical and numerical efforts have been done to understand the dynamics of compressible flows \citep{Z1993,CL2002,Ga2011,B2013,Y2017}. Nearly incompressible (NI) MHD theory is an intermediate model between compressible and incompressible descriptions. Using a particular expansion technique, \citet{Z1993} have derived different NI MHD equations depending on the $\beta$ plasma parameter (ratio between fluid and magnetic pressure). From this NI perspective, one would expect that at high $\beta$ and low Mach number the leading order description would be IMHD~\citep{Sa2007}, with isotropic variances and anisotropic spectra. However, this theoretical predictions are subjected to initial conditions and forcing expressions. In contrast, the low $\beta$ NI MHD theory predicts an anisotropy in both the variances in both the variances and the spatial spectra, which has been observed in the solar wind \citep{Sm2006} and confirmed in numerous simulations \citep[see, e.g.,][]{Ou2016}. \citet{CL2002} presented a theoretical model for CMHD isothermal turbulence in the low $\beta$ regime, and numerically tested it for moderate spatial resolution ($256^3$ grid points). The authors separated the different fluctuation modes and reported different theoretical scalings for each branch, namely an anisotropic Kolmogorov spectrum for the Alfv\'en and slow modes $k_\perp^{-5/3}$ and an isotropic one $k^{-5/3}$ for the fast mode. Using weak turbulence theory \citep{Ku2001}, \citet{Ch2008} also considered the low $\beta$ regime and derived a set of kinetic equations that provide an approximate description of nonlinear processes in collisionless plasmas. Neglecting the slow magnetosonic branch, \citet{Ch2005} used this model to conclude that three-wave interactions transfer energy to high-frequency fast magnetosonic waves and to a lesser extent to high-frequency Alfv\'en waves. The author also predicted a $\sim k^{-7/2}$ power spectra for the fast magnetosonic branch for low $\beta$ values. Direct evidence from direct numerical simulations of CMHD turbulence of the excitation of these waves is still lacking, and thus which energy transfer mechanism is dominant is unclear.

The main objective of the present paper is to study the interplay between the different wave modes in a CMHD developed turbulent regime using the spatio-temporal spectrum \citep{Cl2015a}. This technique allows for direct identification of all wave modes in a turbulent system, and precise quantification of the amount of energy in each mode as a function of the wavenumber. We keep in mind that in strong turbulence, much of the energy resides in modes that are not linear eigenmodes, but rather might be described as zero frequency turbulence. Both low and high $\beta$ regimes and small Mach numbers are considered, situations that are relevant for the solar wind and planetary magnetosheaths. The paper is organized as follows: in Section \ref{section2} we present the CMHD model, where in sub-section \ref{model} we show the set of equations and the normal modes of the CMHD model, in sub-section \ref{setup} we describe the numerical setup used for the study and in sub-section \ref{spatio-tem} we briefly explain the spatio-temporal spectrum technique. In Section \ref{results} we present our results for both low and high $\beta$. Finally, in Section \ref{conclus} we summarize our main findings.

\section{Equations, numerical simulations, and analysis}\label{section2}

\subsection{Compressible MHD equations}\label{model}

The 3D CMHD model is given by the mass continuity equation, the momentum equation, the induction equation for
the magnetic field, and an equation of state for the plasma, which is
assumed here to be polytropic,
\begin{align}\label{modeld:1}
    & \frac{\partial \rho}{\partial
    t}+\boldsymbol\nabla\cdot(\textbf{u}\rho)=0, \\ \label{modeld:2} &
    \frac{\partial \textbf{u}}{\partial t} +
    \textbf{u}\cdot\boldsymbol\nabla\textbf{u}=-\frac{\boldsymbol\nabla
    p}{\rho} + \frac{\textbf{J}\times\textbf{B}}{4\pi\rho} + \nu'
    \bigg[\nabla^2\textbf{u}+\frac{1}{3}\boldsymbol\nabla(
    \boldsymbol\nabla\cdot\textbf{u})\bigg] ,
    \\ \label{modeld:3} & \frac{\partial \textbf{B}}{\partial t} =
    \boldsymbol\nabla\times\left(\textbf{u}\times\textbf{B}\right) +
    \eta' \nabla^2 \textbf{B} , \\ \label{modeld:4} &
    \frac{p}{\rho^{\gamma}}=\text{constant} ,
\end{align}
where \textbf{u} is the fluctuating velocity field, $\textbf{B}=\textbf{B}_0+\textbf{b}$ is the total magnetic field, and $\rho$ is the density. In addition, $\textbf{J}=(4\pi/c)\boldsymbol\nabla\times\textbf{B}$ is the electric current, $p$ the scalar pressure, $\gamma=5/3$ the polytropic index, and $\nu'$ and $\eta'$ are the viscosity and magnetic diffusivity, respectively. {The dissipation terms used in Eqs. \eqref{modeld:2} and \eqref{modeld:3} are not intended to correspond precisely to the dissipation mechanisms in a collisionless plasma. The purpose of these terms is only to dissipate energy at scales smaller than the MHD scales, while allowing us to study with an adequate scale separation compressible effects at the largest scales. Following the usual assumptions done when studying incompressible and compressible (at low Mach numbers) MHD turbulence \citep[see e.g.,][]{Bi1997,CL2002,BC2013}, we take the viscosity and magnetic diffusivity to be independent of the density.}

The set of equations \eqref{modeld:1}-\eqref{modeld:4} can be written in
a dimensionless form in terms of a typical length scale $L_0$ , a
mean scalar density $\rho_0$ and pressure $p_0$, a
typical magnetic field magnitude $b_{rms}$, and a typical velocity field
magnitude $u_{rms}= b_{rms}/\sqrt{4\pi\rho_0}$ (i.e., the r.m.s. Alfv\'en
velocity). The resulting dimensionless equations are
\begin{align}\label{model:1}
    & \frac{\partial \rho}{\partial
    t}+\boldsymbol\nabla\cdot(\textbf{u}\rho)=0, \\ \label{model:2} &
    \frac{\partial \textbf{u}}{\partial t} +
    \textbf{u}\cdot\boldsymbol\nabla\textbf{u}=-\beta\frac{\boldsymbol\nabla
    p}{\rho} + \frac{\textbf{J}\times\textbf{B}}{\rho} + \nu
    \bigg[\nabla^2\textbf{u}+\frac{1}{3}\boldsymbol\nabla(
    \boldsymbol\nabla\cdot\textbf{u})\bigg] ,
    \\ \label{model:3} & \frac{\partial \textbf{B}}{\partial t} =
    \boldsymbol\nabla\times\left(\textbf{u}\times\textbf{B}\right) +
    \eta \nabla^2 \textbf{B} , \\ \label{model:4} &
    \frac{p}{\rho^{\gamma}}=\text{constant} .
\end{align}
Here, $\nu$ and $\eta$ are the dimensionless viscosity and magnetic diffusivity (i.e., the inverse of Reynolds and magnetic Reynolds numbers) respectively, and $\beta\equiv(c_s/u_A)^2$ is the plasma beta, i.e., the ratio of plasma pressure to magnetic pressure, with $c_s=\sqrt{\gamma p_0/\rho_0}$ the sound speed and $u_A= B_0/\sqrt{4\pi\rho_0}$ the Alfv\'en
velocity. The $\beta$ parameter separates two different limiting cases, the magnetic pressure dominated regime ($\beta \ll 1$) and the plasma pressure dominated regime ($\beta \gg 1$).

Linearizing equations \eqref{model:1}-\eqref{model:4} around a static equilibrium (i.e., $\textbf{u}_0=0$) with a homogeneous magnetic field $\textbf{B}_0 = B_0~\hat{z}$, a constant density $\rho_0$, and a constant pressure  $p_0$, we obtain the dispersion relation $\omega(\textbf{k})$ of small amplitude waves propagating in the plasma. As usual, the dispersion relation relates the angular frequency $\omega$ of the waves with its wave vector $\textbf{k}$. It is straightforward \citep[e.g.][]{F2014} to show that there are three independent propagating modes (or waves) for a CMHD plasma, which correspond to the so-called Alfv\'en waves (A), fast (F) and slow (S) magnetosonic waves,
\begin{align}\label{alfven}
    & \omega_A^2(k) = k_\parallel^2 u_A^2 \\ \label{ms} &
    \omega_{F,S}^2(k) =
    k^2 u_A^2\left[\frac{(1+\beta)}{2}\pm\sqrt{\frac{(1+\beta)^2}{4}-\beta
    \bigg(\frac{k_\parallel}{k}\bigg)^2}\right],
\end{align}
where $k_\parallel$ is the wavenumber component along the external magnetic field, and $k=|\textbf{k}|=\sqrt{k_\parallel^2+k_\perp^2}$. Alfv\'en waves are incompressible fluctuations transverse to the magnetic guide field. In the dispersion relation of magnetosonic waves $\omega_{F,S}(k)$, the plus sign on the r.h.s.~of the equation \eqref{ms} corresponds to fast waves, and the minus sign to slow waves. Both fast and slow magnetosonic modes carry density fluctuations, and their magnetic field perturbations have longitudinal and transverse components. Note that for the perpendicular propagation (i.e., $k_\parallel =0$ and $k_\perp \neq 0$) the Alfv\'en and slow modes become non-propagating modes (i.e., $\omega_{A,S}=0$) and are degenerate, but they can be distinguished using their different polarization, since $\delta B_{\parallel, A} = 0$ and $\delta B_{\parallel, S} \neq 0$ (where $\delta B_\parallel$ is the magnetic fluctuations parallel to the guide field). To these non-propagating solutions one must add the entropy mode $\omega_E=0$ characterized by density and entropy fluctuations only. These three non-propagating solutions have their nonlinear counterparts in MHD equilibrium solutions \citep[see, e.g.][]{K1973}, which are likely to develop in turbulent plasmas. As the main goal of the present paper is to identify the various possible waves and structures in the simulations we adopt the assumption that energy that is concentrated closely to the linear dispersion relation can be explained by linear and weak turbulence theories, while any spread round, or away from, those linear curves is a sign of strong turbulence that requires fully nonlinear theories to be understood.

\subsection{Numerical setup}\label{setup}

The 3D CMHD equations \eqref{model:1}-\eqref{model:4} were numerically solved using the Fourier pseudospectral code GHOST \citep{Go2005,Mi2011} with a new module for compressible flows based on previously developed codes \citep{Gh1993,D2005}. The scheme used ensures exact energy conservation for the continuous time spatially discrete equation \citep{Mi2011} (as well as conservation of all other quadratic invariants in the system). The discrete time integration used is a
second-order Runge-Kutta method. Since computation of the
spatio-temporal spectra described below requires a significant amount
of data storage, we used moderate linear spatial resolutions $N=512$
in a 3D periodic box. For simplicity, we used identical dimensionless viscosity and magnetic diffusivity, $\nu=\eta=1\times10^{-3}$ (i.e., the magnetic Prandtl number is $P_m=1$).

The initial state of our simulations corresponds to density, velocity and magnetic fields amplitude fluctuations equal to zero. For all times $t>0$, the velocity field and the magnetic vector potential are forced by a mechanical forcing ${\bf F}$ and electromotive forcing  ${\boldsymbol \epsilon}$, respectively. The mechanical and electromotive forcings are uncorrelated and they inject neither kinetic nor magnetic helicity. At $t=0$, for each forcing function, a random 3D isotropic field ${\bf f}_{\bf k}$ is generated in Fourier space, by filling the components of all modes in a spherical shell with $1 \leq k \leq 2$ with amplitude $f_0$ and a random phase $\phi_{\bf k}$ for each wave vector {\bf k}. Here $k=1$ is refers to the longest wavelength in the periodic box. Then, the Fourier coefficients of a forcing with zero divergence are obtained as,
\begin{equation}\label{rnd}
    {\bf F}_{\bf k} = \frac{{\bf k} \times {\bf f}_{\bf k}}{k}.
\end{equation}
The same process is repeated to generate ${\boldsymbol \epsilon}_{\bf k}$
(note that this satisfies the Coulomb gauge used by the code when evolving the vector potential). An amplitude $f_0 =0.15$ is used for the mechanical and electromotive forcings, and the set of random phases of the two forces are independent. Random phases were also slowly evolving in time, to avoid introducing long-term correlations, but also to prevent introducing very fast time scales. To this end a new set of random phases $\phi_{\bf k}$ is generated for each forcing function every 1/2 turnover time. Finally, the forcings ${\bf F}$ and ${\boldsymbol \epsilon}$ are linearly interpolated from their previous states to the new random states on 1/2 turnover time, and the process is then repeated.

We performed two numerical simulations, both with a weak compressible sonic Mach number $M_s= u_{rms}/c_s=0.25$, but with different values of $B_0$, and thus different values of $\beta$. In one simulation we used a strong guide magnetic field $B_0=8$, which corresponds to $\beta=0.25$. In the other simulation we used a moderated guide field $B_0=2$, which
corresponds to $\beta=4$. This allowed us to investigate two different regimes, i.e., the magnetic and plasma pressure dominated regimes. Note however that modifying the guide field magnitude results as well in the modification of the nonlinearity parameter $\chi$ (defined above for IMHD turbulence). The simulation with $\beta = 4$ corresponds to a nonlinearity parameter (at the driving scale) that is four times higher than the case at $\beta=0.25$. We will return to this point in the discussion Section.

\subsection{Spatio-temporal spectrum}\label{spatio-tem}

The spatio-temporal spectrum allows identification of waves in turbulent flow. The technique consists of calculating the complete spectrum in wavenumber and frequency for all available Fourier modes in a numerical simulation or an experiment \citep{S2003,Cl2015a}. As a result, it can separate between modes that satisfy a given dispersion relation (and are thus associated with waves) from those associated to nonlinear structures or turbulent eddies, and quantify the amount of energy carried by each of them. The method we use does not require the pre-existence of wave modes or eddies. Quantifying the relative importance of each of them and understanding the physics that controls it is the main outcome expected from the present analysis. In the following, the spatio-temporal magnetic energy spectral density tensor is defined as
\begin{equation}
    E_{ij}(\textbf{k},\omega) =
    \frac{1}{2}\hat{B}_i^{*}(\textbf{k},\omega)\hat{B}_j
    (\textbf{k},\omega) ,
\end{equation}
where $\hat{B}_i(\textbf{k},\omega)$ is the Fourier transform in space and time of the $i$-component of the magnetic field
$\textbf{B}(\textbf{x},t)$ and where the asterisk implies the complex conjugate. The magnetic energy is associated with the trace of $E_{ij}(\textbf{k},\omega)$.

As the external magnetic field ${\bf B}_0$ in the simulations points in $\hat{z}$, in practice we will consider either $i=j=y$ or $i=j=z$, to identify different waves based on their polarization (either transverse or longitudinal with respect to the guide field). It is worth mentioning that spatio-temporal spectra have been used before in numerical simulations and experiments of rotating turbulence \citep{Cl2014}, stratified turbulence \citep{Cl2015c}, quantum turbulence \citep{Cl2015b}, and IMHD turbulence simulations \citep{Me2015,Me2016,Lu2016} and in spacecraft observations~\citep{S2003,S2010}. In the present paper we use the technique to investigate the interplay between Alfv\'en and magnetosonic waves in CMHD turbulence.

In all cases, the temporal extent of the data used to calculate the spatio-temporal spectra was longer than at least one period of the slowest wave in the system, and the temporal data cadence was at least twice as fast as the fastest wave. The emergence of fluctuations occurring on very long time scales, corresponding to $1/f$ noise in the power frequency spectrum, have been observed in systems such as IMHD with a background magnetic field or in rotating fluid turbulence \citep{Ma1986,V1989,M2007,D2011,D2014,H2015}. However, in the present paper we emphasize the wave modes at higher frequencies and not the dominance by $1/f$ noise at long time scales.

\begin{figure}
    \centering
    \includegraphics[width=.45\textwidth]{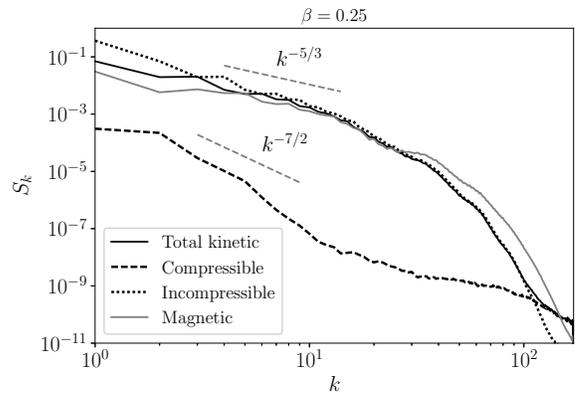}
    \caption{Spatial spectrum $S_k$ of the total magnetic and kinetic
    energy (in solid gray and black lines, respectively). The dotted
    and dashed lines correspond to the kinetic energy spectra of the
    incompressible and compressible components of the flow,
    respectively. Two scaling laws, $\sim k^{-5/3}$ and $\sim
    k^{-7/2}$ are shown as references.}
    \label{1spatio}
\end{figure}

\begin{figure}
    \centering
    \includegraphics[width=.5\textwidth,center]{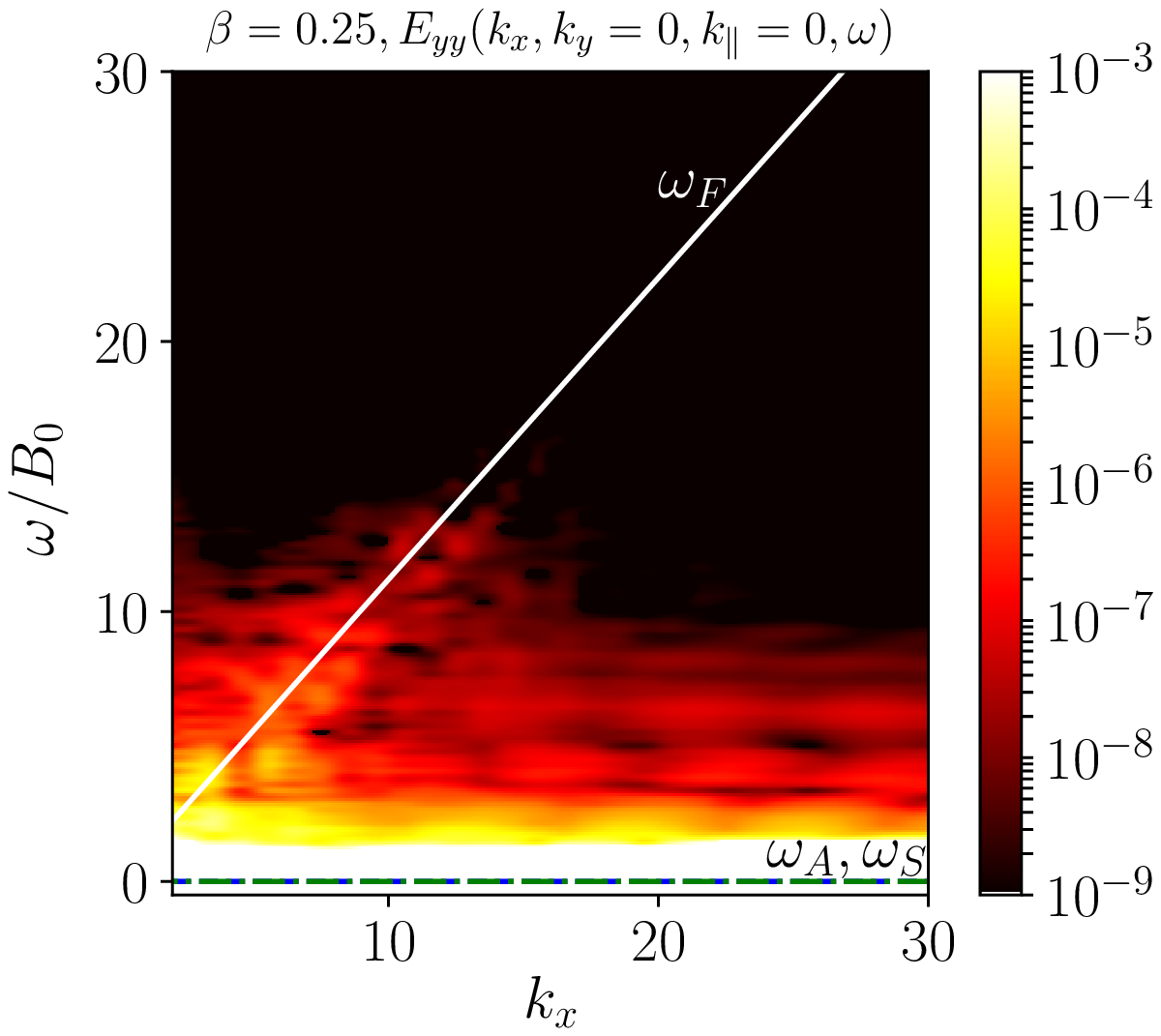}
    \vspace{-.8cm}
    \caption{({\it Color online}) Spatio-temporal spectrum
    $E_{yy}(k_x,k_y=0,k_ \parallel=0,\omega)$  for the magnetic field
    fluctuations perpendicular to \textbf{B}$_0$,
    for $\beta=0.25$. The dashed, solid, and dash-dotted lines
    correspond to the linear dispersion relationd of Alfv\'en waves
    $\omega_A$, of fast magnetosonic waves $\omega_F$, and of slow
    magnetosonic waves $\omega_S$, respectively.}
    \label{1perpB}
\end{figure}

\begin{figure}
    \centering
    \includegraphics[width=.5\textwidth,center]{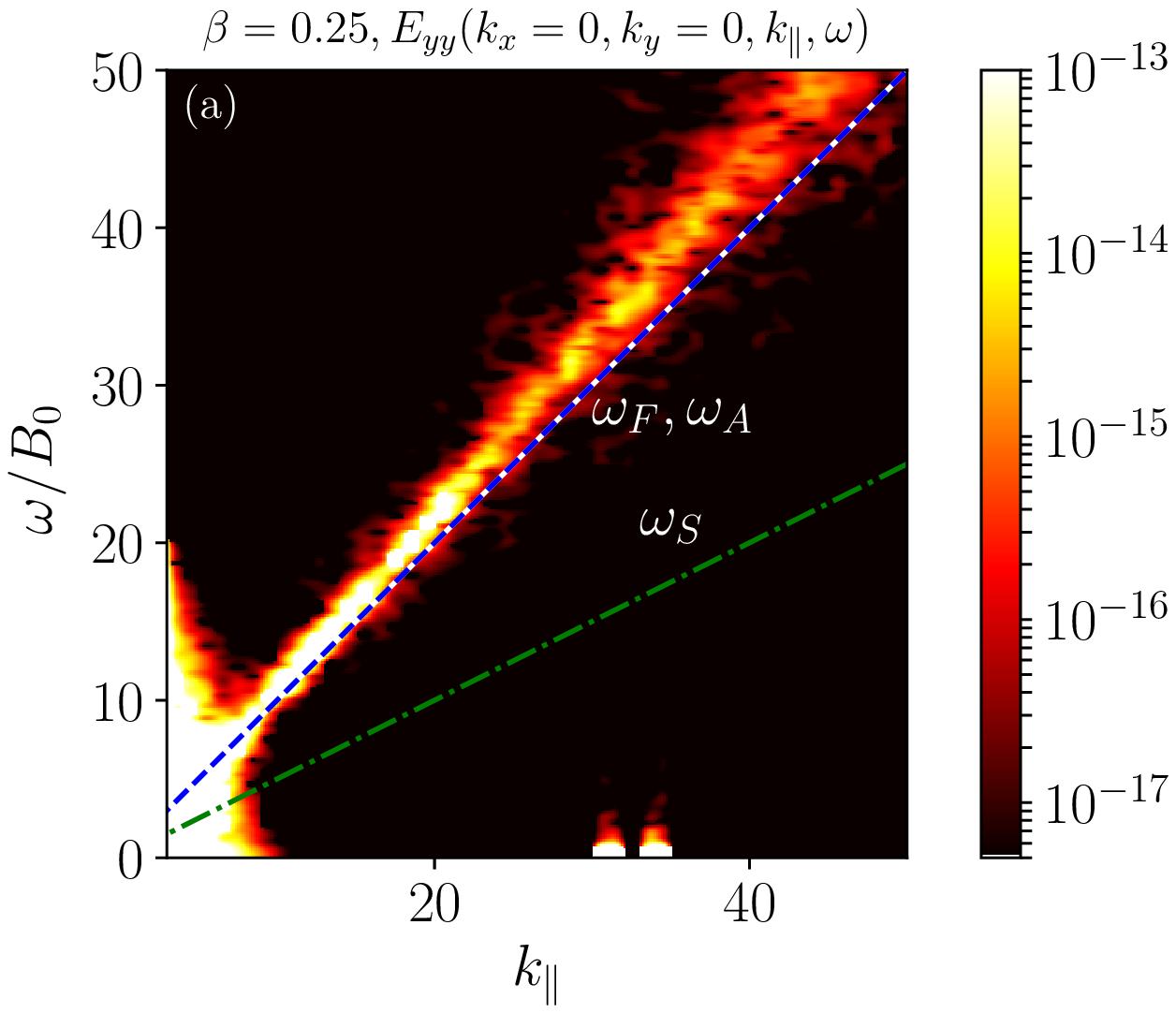}
    \includegraphics[width=.5\textwidth,center]{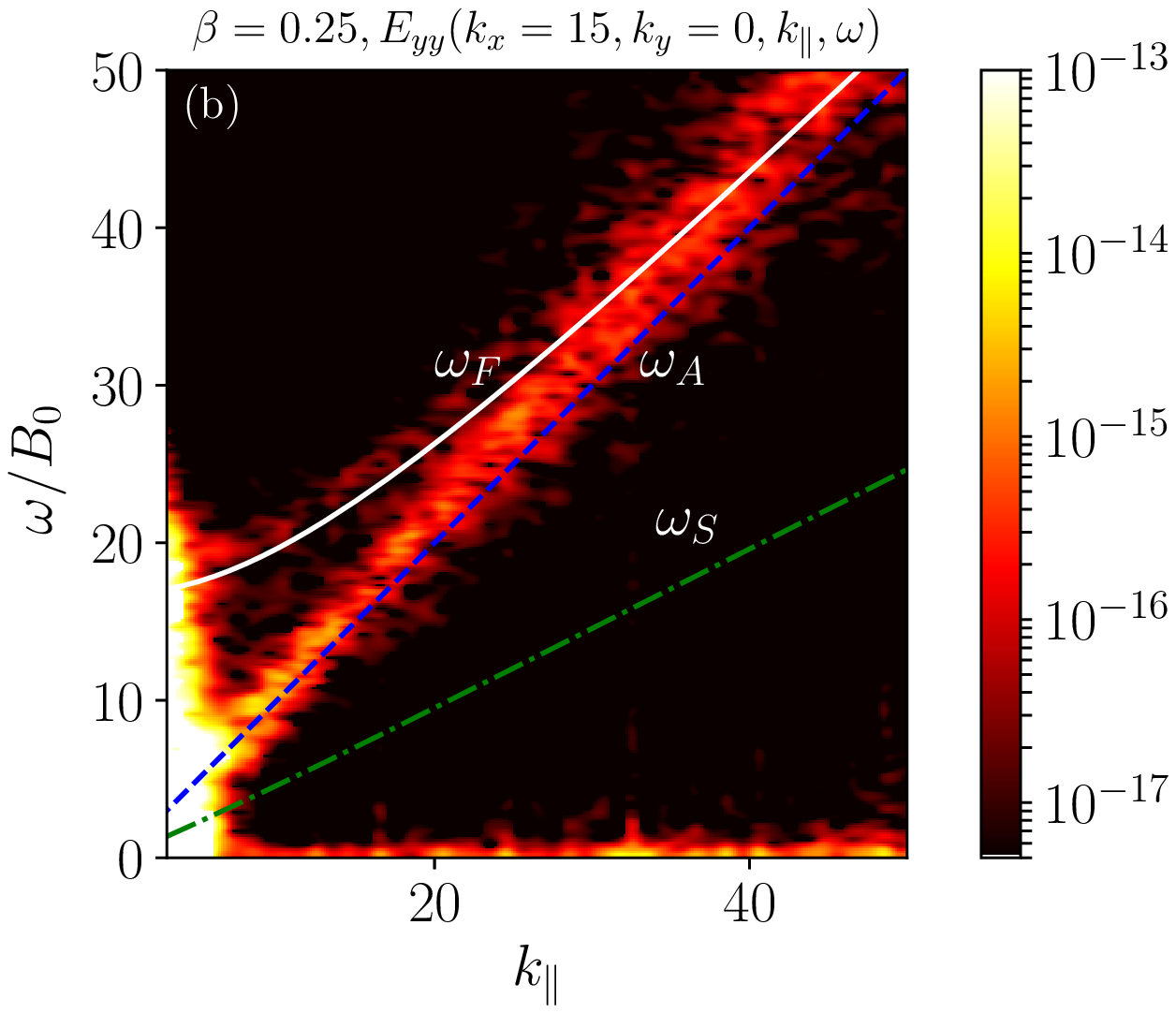}
    \vspace{-.8cm}
    \caption{({\it Color online}) Spatio-temporal spectrum
    $E_{yy}(k_x=0,k_y,k_\parallel,\omega)$ for the magnetic field
    fluctuations perpendicular to \textbf{B}$_0$, for
    $\beta=0.25$. The spectrum is shown as a function of $\omega$ and
    $k_\parallel$ for fixed $k_y=0$ (a) and $k_y=15$
    (b). The dashed, solid, and dash-dotted lines
    correspond to the linear dispersion relationd of Alfv\'en waves
    $\omega_A$, of fast magnetosonic waves $\omega_F$, and of slow
    magnetosonic waves $\omega_S$, respectively. For $k_\perp=0$ the
    Aflv\'en and fast branches coincide.}
    \label{1perp}
\end{figure}

\section{Numerical Results and Discussion}
\label{results}

\subsection{Low $\beta$ regime}
\label{low}

Reduced spatial spectra are obtained from the general spatio-temporal spectra by integration over all frequencies and over all wave vectors in spherical shells of radius $k$. As an example, for the magnetic energy the spatial isotropic (omnidirectional) spectrum satisfies,
\begin{equation}
S_k(k) = \sum_\omega \sum_{k \le |{\bf k}| < k+1}
  \left[ E_{xx}({\bf k},\omega) + E_{yy}({\bf k},\omega) + E_{zz}({\bf
      k},\omega) \right] .
\end{equation}
Similarly, we computed the spatial isotropic spectrum for the kinetic energy. Besides, we computed the compressible and incompressible kinetic spectrum of the flow using the usual Helmholtz decomposition \citep[see, e.g.][]{Da2006}. In Fig.~\ref{1spatio} we show the spatial energy spectra $S_k$ of the kinetic and the magnetic energy for the simulation with $\beta=0.25$. We also show the power spectra of the compressible and incompressible components of the velocity field.

An inertial range compatible with a $\sim k^{-5/3}$ can be observed in Fig.~\ref{1spatio} for the total kinetic energy, the incompressible kinetic energy, and the magnetic energy. The compressible kinetic energy spectrum is weaker and steeper; the $\sim k^{-7/2}$ scaling predicted by \citet{Ch2005} is shown for reference. A detailed study of these scaling laws would require larger spatial resolutions, which are outside the scope of this work. Note also that while the vast majority of the kinetic energy is located in its incompressible component, it is known that the small compressible component can still affect the flow dynamics in this regime. For example, direct numerical simulations performed with the same Mach and $\beta$ numbers show that proton acceleration is significantly enhanced when compared to the incompressible case \cite{G2016}.

Spatial analysis alone cannot fully determine the presence of Alfv\'en or magnetosonic waves, much less determine which (if any) dominates the dynamics; to do this we must turn to spatio-temporal analysis. Fig. \ref{1perpB} shows the spatio-temporal spectrum of the perpendicular magnetic field fluctuations $E_{yy}(k_x,k_y=0,k_\parallel=0,\omega)$ for fixed $k_y=k_\parallel=0$ for the same simulation as in Fig.~\ref{1spatio} (since the spatio-temporal spectrum is four dimensional, we fix two components of ${\bf k}$ to plot the remaining component against the frequency). The dispersion relations for Alfv\'en and magnetosonic waves given by equations \eqref{alfven} and \eqref{ms} are shown in dashed, dash-dotted, and dotted lines, respectively. The energy accumulates mainly for low $\omega/B_0$ ($\lesssim 4$) and $k_\parallel=0$, i.e., in two-dimensional (2D) modes, which correspond to turbulent eddies and which is to be expected for IMHD turbulence with a guide field. Fig.~\ref{1perp} shows the spatio-temporal spectrum of the perpendicular magnetic field fluctuations $E_{yy}(k_x=\nobreak 0,k_y,k_\parallel,\omega)$, for fixed values of $k_y = 0$ (Fig.~\ref{1perp}(a)) and $k_y = 15$ (Fig.~\ref{1perp}(b)). In this case, energy accumulates mostly in modes with low $k_\parallel$ and low $\omega/B_0$, typically $k_\parallel \lesssim 5$ and $\omega/B_0 \lesssim 20$. {Fig. \ref{1perp}(a) shows the presence of energy at around $k_\parallel=32$ and $\omega \approx 0$. These two peaks are again repeated at $k_\parallel=64, 96$ and 128. They are probably due to finite size and/or finite sampling time effects but their overall energy is small and they do not seem to play any significant role in the turbulent and wave dynamics of the system.} Fig. \ref{1perpB} and Fig. \ref{1perp} are compatible with the NI MHD theory for low $\beta$, where the leading order description is two-dimensional with compressible corrections. Thus, as in the case of IMHD turbulence \cite{A2011,S2016}, the presence of a strong guide field produces strong bidimensional components even in the presence of weak compressibility.

\begin{figure}
    \centering
    \includegraphics[width=.5\textwidth,center]{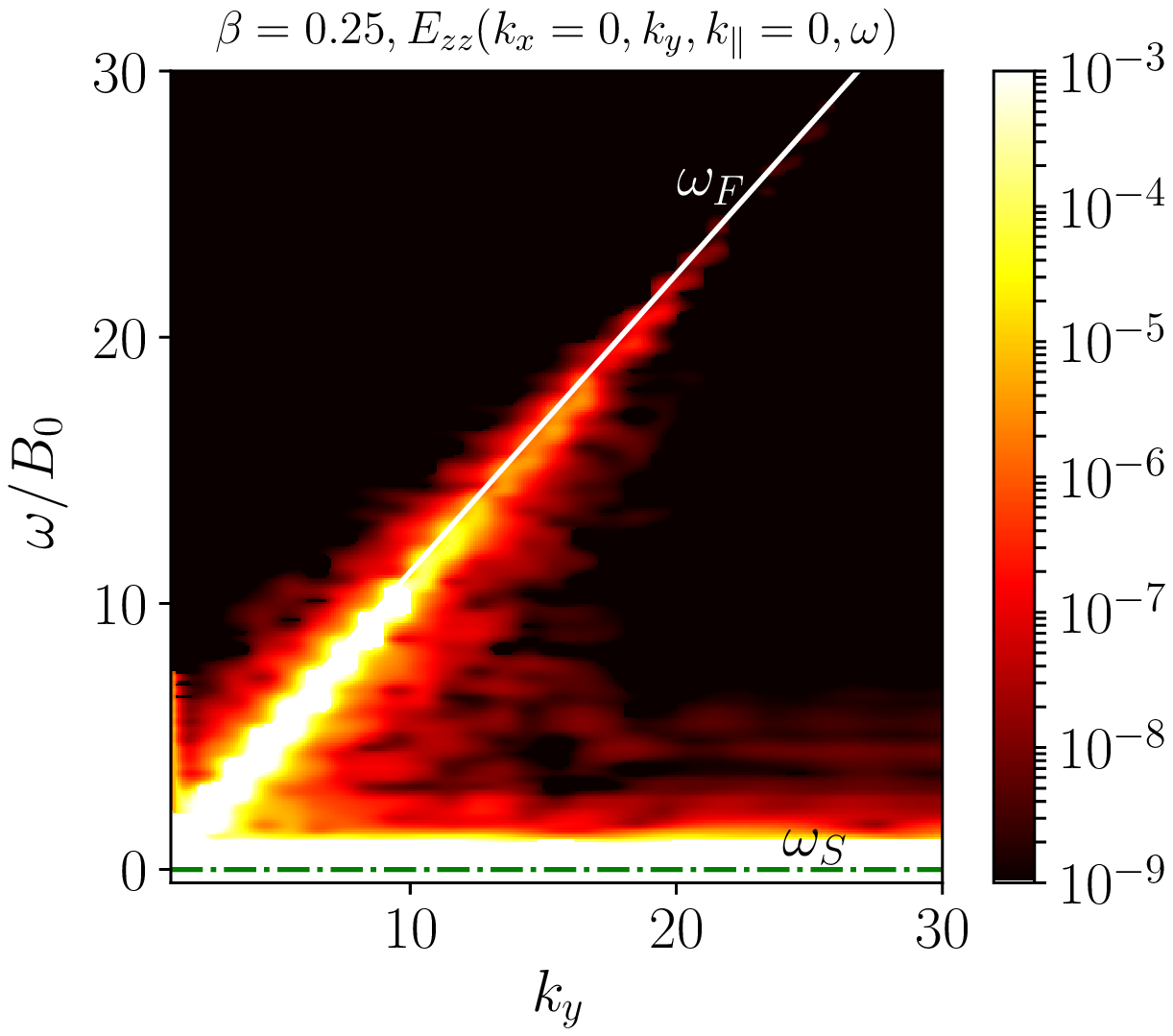}
    \vspace{-.8cm}
    \caption{({\it Color online}) Spatio-temporal spectra
    $E_{zz}(k_x=0,k_y,k_\parallel=0,\omega)$ for the magnetic field
    fluctuations parallel to \textbf{B}$_0$, for $\beta=0.25$.
    The solid and dash-dotted lines correspond to the linear
    dispersion relationd of of fast magnetosonic waves $\omega_F$,
    and of slow magnetosonic waves $\omega_S$, respectively.}
    \label{1para}
\end{figure}

For modes with $k_\parallel \gtrsim5$, energy in Fig.~\ref{1perp}(a) then accumulates around the Alfv\'en wave branch $\omega_A$ (note that for $k_\perp=0$, the Alfv\'en and fast branches overlap), while Fig.~\ref{1perp}(b) some energy also present in the vicinity of the fast magnetosonic branch $\omega_F$ and along $\omega =0$. Both Figures do not show energy spread along the slow magnetosonic branch $\omega_S$. In other words, energy in high frequency modes ($\omega>0$ and with $k_\parallel \gtrsim 5$) accumulates near the  dispersion relation of the fastest waves, in agreement with predictions from weak turbulence \cite{Ch2005,Ch2008}. At high parallel wavenumbers energy accumulation deviates slightly from the linear dispersion relations, but is still concentrated around specific modes, indicating possible coupling of fast magnetosonic and Alfv\'en waves, or nonlinear corrections to the dispersion relations. In contrast with what was previously suggested \cite{CL2002}, fast magnetosonic waves are not suppressed by Alfv\'en waves, but they do not dominate the dynamics either as predicted using weak wave turbulence theory \cite{Ch2005,Ch2008}.

Fast magnetosonic waves can be separated from the Alfv\'en waves by looking at the spatio-temporal spectrum of parallel magnetic field fluctuations $E_{zz}(k_x=\nobreak0,k_y,k_\parallel=0,\omega)$, shown in Fig.~\ref{1para}. The Alfv\'en waves do not contribute to the parallel component of the magnetic field energy since their magnetic perturbations are perpendicular to the guide field. In Fig.~\ref{1para} energy accumulates in two regions: at high frequency near the fast magnetosonic branch, and at low frequency near $\omega=0$ modes. Note that the spread around the linear dispersion relations curves is likely to be caused by nonlinear effects. It is worth noticing that, unlike in Fig.~\ref{1perp}(a), energy in Fig.~\ref{1para} does not show any shift toward higher frequency than the linear dispersion relation of the fast mode.

\begin{figure}
    \centering
    \includegraphics[width=.45\textwidth]{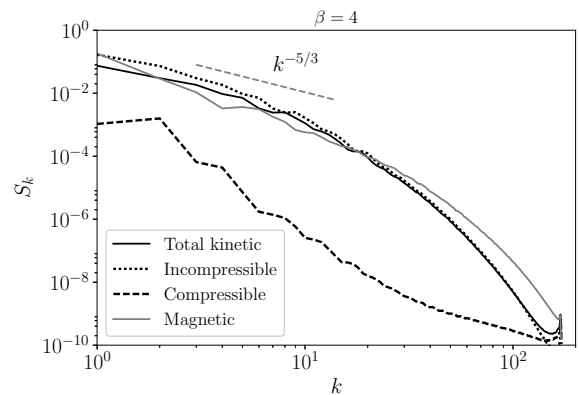}
    \caption{Spatial spectrum $S_k$ of the total magnetic and kinetic
    energy (in solid gray and black lines, respectively). The dotted
    and dashed lines correspond to the power spectra of incompressible
    and compressible components of the flow, respectively. A
    $\sim k^{-5/3}$ scaling is shown as reference.}
    \label{2spatio}
\end{figure}

\begin{figure}
    \centering
    \includegraphics[width=.5\textwidth,center]{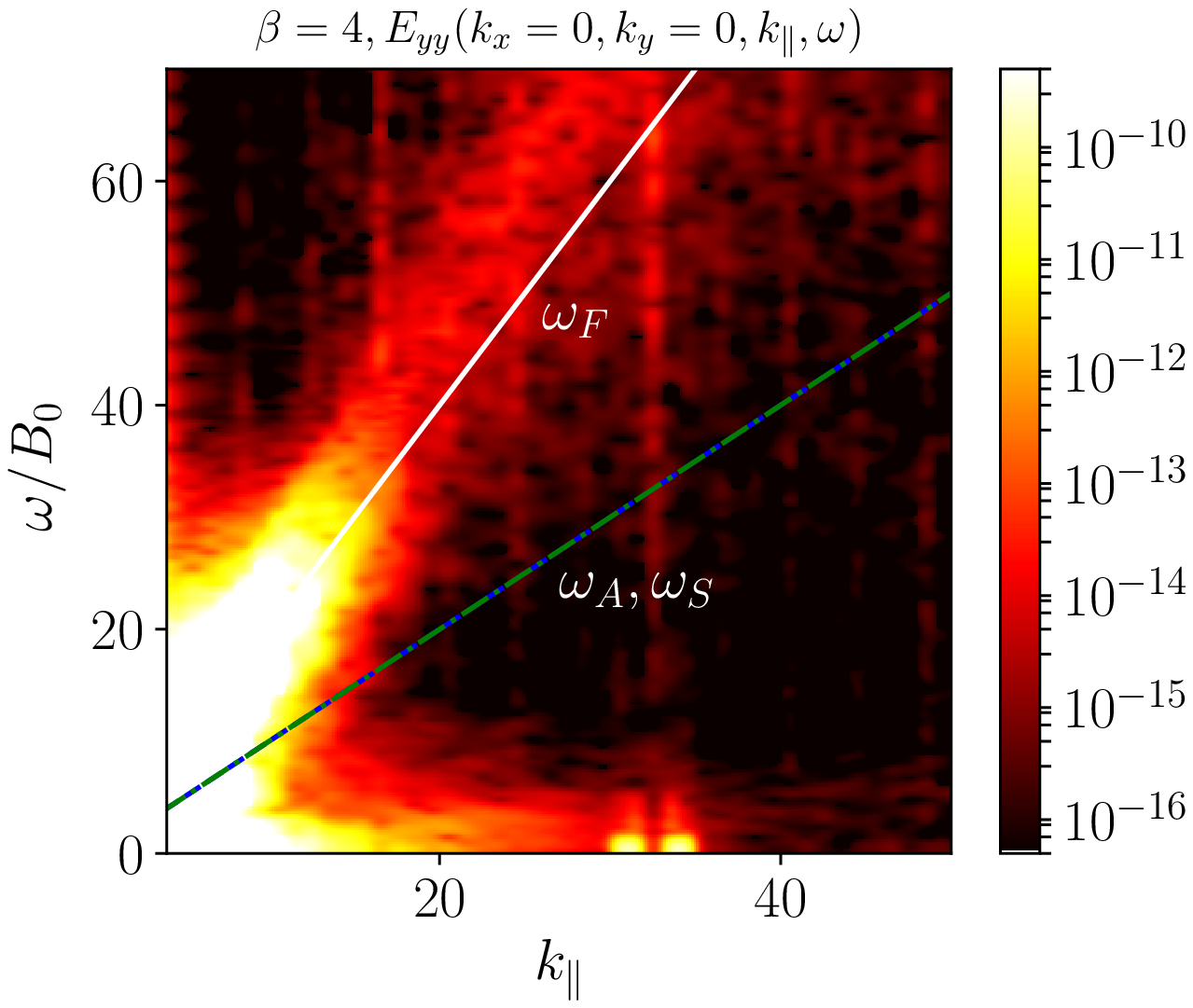}
    \vspace{-.3cm}
    \caption{({\it Color online}) Spatio-temporal spectrum
    $E_{yy}(k_x=0,k_y=0,k_\parallel,\omega)$ of the magnetic field fluctuations perpendicular to \textbf{B}$_0$, for the run with $\beta=4$. The dashed, solid, and dash-dotted lines correspond to the linear dispersion relation of Alfv\'en waves $\omega_A$, of fast magnetosonic waves $\omega_F$, and of slow magnetsonic waves $\omega_S$,  respectively (in this case, for $k_\perp=0$ the dispersion relations of slow and Alfv\'en waves coincide).}
    \label{2perp}
\end{figure}

\begin{figure}
    \centering
    \includegraphics[width=.5\textwidth,center]{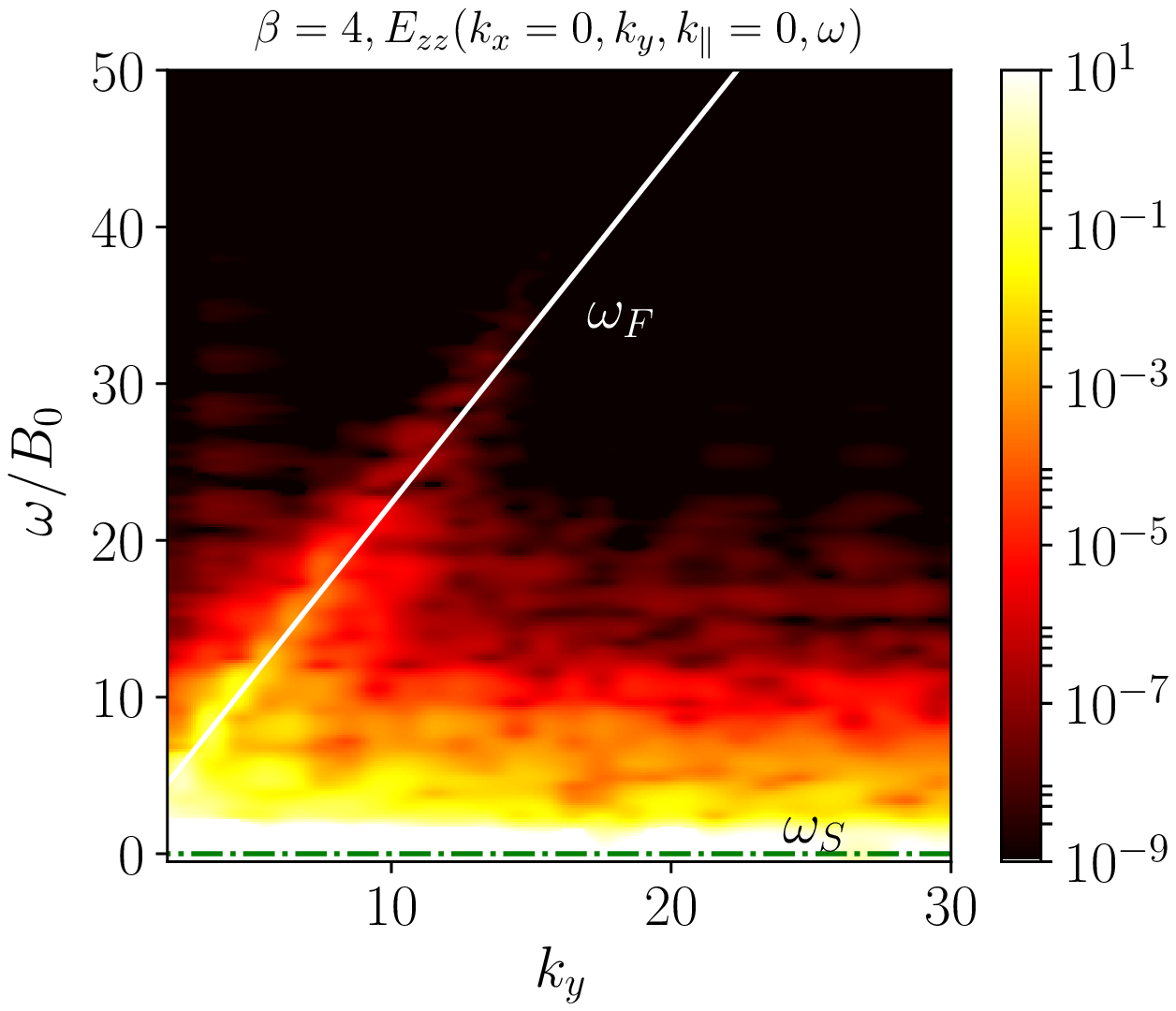}
    \vspace{-.4cm}
    \caption{({\it Color online}) Spatio-temporal spectrum
    $E_{zz}(k_x=0,k_y,k_\parallel=0,\omega)$ of the magnetic field
    fluctuations parallel to \textbf{B}$_0$, for $\beta=4$. The
    solid and dash-dotted lines correspond to the linear dispersion
    relation of fast magnetosonic waves $\omega_F$, and of slow
    magnetsonic waves $\omega_S$,  respectively}
    \label{2para}
\end{figure}

\subsection{High $\beta$ regime}
\label{high}

In Fig.~\ref{2spatio} we show the spatial energy spectra $S_k$ of the kinetic and the magnetic energy for the simulation with $\beta=4$; we also show the energy spectra of the compressible and incompressible components of the velocity field. An inertial range roughly compatible with $\sim k^{-5/3}$ is observed for the total kinetic energy, the incompressible kinetic energy, and the magnetic energy. No discernible scaling is present in the compressible kinetic energy spectrum. Once again, the vast majority of the kinetic energy is in the incompressible component of the flow.

To determine the presence of waves in the higher frequency part of the  turbulent flow we turn once more to the spatio-temporal spectrum. Fig.~\ref{2perp} and Fig.~\ref{2para} show the spatio-temporal spectrum of the perpendicular magnetic field fluctuations $E_{yy}(k_x=0,k_y=0,k_\parallel,\omega)$, and the spectrum of parallel magnetic field fluctuations $E_{zz}(k_x=0,k_y,k_\parallel=0,\omega)$, respectively. The dispersion relations given by equations~\eqref{alfven}-\eqref{ms} are in dashed, dash-dotted, and dotted lines. Fig.~\ref{2perp} shows that for frequencies $\omega/B_0 \gtrsim 10$ and $k_\parallel \gtrsim 5$, the only wave modes that are now excited are the fast magnetosonic ones, and no apparent traces of Alfv\'en waves, which coincide with the slow mode in this case. Fig.~\ref{2perp} shows also a smaller amount of energy near the $\omega=0$ (and $k_\parallel \neq 0$) modes. Fig.~\ref{2para} shows that most of the energy lies along the slow mode $\omega_S$ curve, and a smaller fraction of energy follow the fast mode curve $\omega_F$. The spread around those curves is likely to be due to stronger nonlinear interactions that can generate 2D structures that can coincide with the curve $\omega_S$ in Fig.~\ref{2para}.

It is worth mentioning that, for an IMHD run with a guide field ${\bf B}_0$ = 2 (not show here) we obtain a similar result to Fig. \ref{2para}, without the fast magnetosonic trace. This supports the dominance of the 2D (incompressible) structures in Fig.~\ref{2para} rather than the non-propagating (compressible) mode $\omega_S$. Furthermore, the absence of Alfv\'en waves might be due to the weak magnetic guide field used (${\bf B}_0$ = 2), as already was found in IMHD simulations \cite{L2016}. Therefore, in the high $\beta$ regime, fast magnetosonic modes dominate the dynamics (over the Alfv\'en waves) at high frequencies and wavenumbers. However, as we mentioned above, we emphasize that the system in its entirety is dominated mainly by the contributions of 2D modes related to turbulent eddies and non-propagating slow (or entropy) modes ($k_\parallel=0$ and $\omega=0$).

Despite the fact that fast magnetosonic waves concentrate most of the energy in the waves at high frequency, their contribution to the total energy in the system is bounded by the small fraction of energy in compressible motion (see Fig.~\ref{2spatio}). This result is in agreement with recent 3D Landau-fluid simulations~\citep{K2017}. The reason for observing fast modes (rather than Alfv\'en modes) in the high plasma $\beta$ regime remains unclear. We speculate that they might have been favored by the isotropic forcing used in our simulations (the fast modes being the only isotropic modes~\citep{CL2002}). Future numerical simulations with a different (anisotropic) forcing will be needed be needed to unambiguously answer this question.

Another question is whether the different results obtained in the low and high plasma $\beta$ regimes are actually due to the change in the plasma $\beta$ or to that of the nonlinear parameter $\chi$ discussed above, as demonstrated in recent Landau-fluid simulations~\citep{Su2016,K2017}. The fact that the simulation in the low $\beta$ case corresponds to a nonlinearity parameter that is four time smaller than that of the high $\beta$ case may explain the observations of different branches of the linear modes in the former case. It may also explain the broad accumulation of energy around the fast magnetosonic waves in the case of high $\beta$ (Fig.~\ref{2perp}) in comparison to that around Alfv\'en waves in the low $\beta$ case (Fig.~\ref{1para}). However, more numerical simulations are required to answer this question.

\section{Conclusions}\label{conclus}

We used spatio-temporal spectra of different magnetic field components to study waves in compressible MHD turbulence at low and high $\beta$ regimes. In the magnetic pressure dominated regime, we showed direct evidence of the presence of fast magnetosonic and Alfv\'en waves. In particular, we found wavenumber scaling for the spatial spectra compatible with theoretical predictions. We also found that the energy transfer is dominated by the Alfv\'enic or the incompressible fluctuations, and to a lesser extent by fast magnetosonic fluctuations (specially in the perpendicular direction). Although the role of magnetosonic waves is not as important as predicted by some weak wave turbulence theories \citep{Ch2005,Ch2008}, they are not negligible. Moreover, the results confirm that the fastest waves in the system concentrate a non-negligible fraction of the energy at high frequency (even for moderate values of the sonic Mach number), and can thus have a role in the dynamics, with implications for particle acceleration and other processes in the solar wind.

In the high $\beta$ regime, at high frequency only fast magnetosonic waves were present, with no clear trace of Alfv\'en waves. At low frequency, 2D turbulent eddies and non-propagating slow (or entropy) modes may co-exist and seem to carry most of the turbulent energy. This regime is thus similar to that of IMHD with a weak magnetic guide field. The questions as to how the dynamics changes when increasing the magnetic guide field and the Mach number, or when fixing the same plasma $\beta$ and modifying the $\chi$ parameter at the driving scale will be addressed in future studies.
\\

\section*{Acknowledgments}
N.A. is supported through an \'Ecole Polytechnique Postdoctoral Fellowship and by LABEX Plas@Par through a grant managed by the Agence Nationale de la Recherche (ANR), as part of the program “Investissements d’Avenir” under the reference ANR-11-IDEX-0004–02. F.S. and N.A. acknowledge financial support from Programme National Soleil-Terre (PNST). P.C., P.D.M. and P.D. acknowledge support from UBACYT Grant No.~20020130100738BA, and PICT  Grants Nos.~2011-1529 and 2015-3530. W.H.M. is supported in part by the NASA Heliophysics Grand Challenge (NNX14AI63G), LWS (NNX14AI63G) and Heliophysics GI  (NNX17AB79G) programs. P.C.D.L. acknowledges funding from the European Research Council under the European Community’s Seventh Framework Program, ERC Grant Agreement No. 339032.

\bibliographystyle{apsrev4-1}

\end{document}